\documentstyle[aps,preprint,epsf]{revtex}
\draft
\def\cc#1{\mbox{\boldmath${#1}$}}
\input amssymb_simm.sty
\begin{document}
\preprint{\today}
\title{Charge Fluctuations in the Two-Dimensional \\ One-Component Plasma} 
\author{D. Levesque and J.-J. Weis}
\address{Laboratoire de Physique Th{\'e}orique,
  B{\^a}t. 210, Universit{\'e} de Paris-Sud \\ 91405 Orsay Cedex,
  France}
\author{J. L. Lebowitz}
\address{Department of Mathematics and Physics,
  Rutgers University \\ New Brunswick, New Jersey 08903
  }

\date{\today}


\maketitle

\vskip.2truein
\centerline{Dedicated to our friend and colleague George Stell.}

\begin{abstract}
We study, via computer simulations, the fluctuations in the net electric
charge, in a two dimensional one component plasma (OCP) with uniform
background charge density $-e \rho$, in a region $\Lambda$ inside a much
larger overall neutral system.  Setting $e=1$ this is the same as the
fluctuations in $N_\Lambda$, the number of mobile particles of charge $e$.
As expected the distribution of $ N_\Lambda$ has, for large $\Lambda$, a
Gaussian form with a variance which grows only as ${\hat \kappa} |\partial
\Lambda|$, where $|\partial \Lambda|$ is the length of the perimeter of
$\Lambda$.  The properties of this system depend only on the coupling
parameter $\Gamma = kT$ which is the same as the reciprocal temperature in
our units.  Our simulations show that when the coupling parameter $\Gamma$
increases, $\hat \kappa(\Gamma)$ decreases to an asymptotic value ${\hat
\kappa}(\infty) \sim {\hat \kappa}(2)/2$ which is equal (or very close) to
that obtained for the corresponding variance of particles on a rigid
triangular lattice.  Thus, for large $\Gamma$, the characteristic length
$\xi_L = 2 {\hat \kappa}/\rho$ associated with charge fluctuations behaves
very differently from that of the Debye length, $\xi_D \sim 1/\sqrt
\Gamma$, which it approaches as $\Gamma \to 0$.  The pair correlation
function of the OCP is also studied.

\end{abstract}


\bigskip

\narrowtext

\noindent {\bf I.  INTRODUCTION} 

A striking manifestation of the special long range nature of the Coulomb
interactions and the resulting screening \cite{STELL}, \cite{SL} is that
the fluctuations of the net electrical charge $Q_{\Lambda}$, contained in a
subregion $\Lambda$ of a spatially homogeneous and overall neutral
equilibrium macroscopic system, grow only as the surface area $|\partial
\Lambda|$ and not as its volume $|\Lambda|$, the normal behavior of
fluctuations of extensive variables (at critical points the growth is even
faster) \cite{GLM}-\cite{BF}.  This behavior of the charge fluctuations can
be readily understood by considering the (truncated) charge-charge
correlation function $S$ in an overall neutral translation invariant
Coulomb system.  This is given by,
\begin{equation}
\label{I2}
S({\cc r} - {\cc r}^\prime) = \sum_{\alpha, \gamma} e_{\alpha} e_{\gamma}
< \rho_{\alpha}({\cc r}) \rho_{\gamma}( {\cc r}^\prime) > 
\end{equation}
where $\rho_{\alpha}({\cc r})$ the microscopic density of particles of
species $ \alpha$ and $e_{\alpha}$ their charge.
The charge fluctuations of $Q_\Lambda$ is then expressed in terms of
$S$ as 
\begin{equation}
\label{I1}
<Q_{\Lambda}^2> = \int_{\Lambda} d {\cc r} \int_{\Lambda} d {\cc
r}^{\prime} S({\cc r} - {\cc r}^{\prime})
\end{equation}
By introducing the characteristic function $\chi_{\Lambda}(\cc r)$ of
the domain $\Lambda$,

\begin{equation}
\chi_{\Lambda}(\cc r) = \left\{  \begin{array}{cc}
                        1,  & \quad {\cc r} \in \Lambda \\
                        0,  & \quad  {\cc r} \notin \Lambda
                               \end{array} \right.
\end{equation}
the integration in (\ref{I1}) can be extended to all space 
\begin{equation}
\label{I3}
<Q_{\Lambda}^2> = |\Lambda| \int_{{\mathbb{R}}^d} d {\cc r} S(\cc r)
                  - \int_{{\mathbb{R}}^d} d {\cc r} S(\cc r)
\alpha_{\Lambda} (\cc r)
\end{equation}
where 
\begin{equation}
\alpha_{\Lambda} (\cc r) = \int_{{\mathbb{R}}^d} d {\cc r}^\prime
\chi_{\Lambda} ( \cc r + {\cc r}^\prime) \big[ 1-\chi_{\Lambda} ( {\cc
r}^\prime) \big]
\end{equation}

One observes now that
the first term in (\ref{I3}), which is proportional to the volume, vanishes
when the integral of $S$ vanishes, i.e.\ if there is 
``perfect screening'' of the  charges, \cite{GLM}-\cite{GLM1}. Under these
conditions  the only  contribution to (\ref{I3}) comes 
 from the second term which yields, 
when the limit $|\Lambda| \rightarrow
{{\mathbb{R}}^d}$ is taken in a self-similar way,

\begin{equation}
\label{I4}
\lim_{|\Lambda| \to \infty} \frac{ < Q_{\Lambda}^2 >}{| \partial 
\Lambda|}
= -\alpha_d \omega_d \int_0^\infty r^d S(r) dr = ({1 \over 2}\sum_\alpha 
\rho_\alpha e^2_\alpha)\xi_L
\end{equation}
where $r=|\cc r|$, $\omega_d$ is the surface area of a unit sphere in
$d$-dimensions $(\omega_1 = 2)$, and it has been assumed that the system is
also rotationally invariant.  The 
geometrical constant $\alpha_d$ is: $\alpha_3 = 1/4$, $\alpha_2 =
\pi^{-1}$, $\alpha_1 = 1/4$, \cite{GLM}.   The constant $\xi_L$ defines a
characteristic length for the ``charge separation'' associated with the
charge fluctuations \cite{GLM}-\cite{BF}.

We thus see that perfect screening and existence of the integral in
(\ref{I4}) imply surface growth of the variance of charge fluctuations.
The latter, but not the former, will generally be violated when a charged
system with $d$-dimensional Coulomb interactions is confined to a lower
dimensional space, in which case the fluctuations will grow as $|\partial
\Lambda| \log|\Lambda|$, e.g.\ for an $r^{-1}$ potential in $d=2$ or
logarithmic Coulomb potential in $d=1$ \cite{Jancovici}, \cite{MehCos}.  We 
also 
note that starting with a system which is not translation or
rotation invariant, e.g.\ one with periodic structure, we can still apply
(\ref{I3}) and, when perfect screening holds, also (\ref{I4}) after
averaging over translations and rotations.

There are various arguments for expecting perfect screening, also referred
to as the zeroth or charge ``sum rule'', in equilibrium Coulomb systems
\cite{SL} - \cite{GLM1}.  They all involve assumptions about some minimal
decay of correlations in such systems.  This has 
been proven rigorously for classical systems at sufficiently high
temperatures and low densities, i.e.\ in the Debye-H{\"u}ckel regime, where
the decay is exponential, \cite{BRF}, \cite{I}.  For quantum Coulomb
systems the decay is only polynomial but still good enough.  In 
fact one expects that perfect screening will always hold and so $\langle
Q^2_\Lambda \rangle/ |\Lambda| \to 0$ as $|\Lambda| \to \infty$.  Of course
in order to treat particles with charges of different signs via classical
statistical mechanics it is necessary to modify their Coulomb potential at
short range, e.g., by introducing hard cores to prevent collapse.  This is
so in dimensions $d \geq 3$ at all temperatures and in $d=2$ at low
temperatures.  There is no such requirement for quantum systems as long as
either the negative or positive charges (or both) obey Fermi statistics
\cite{LL}.

\noindent {\bf II.  PARTICLE FLUCTUATIONS IN THE OCP}

A particularly interesting example of a system with reduced charge
fluctuations is the one component plasma (OCP).  In this system, used to
model diverse physical situations, one kind of charges, say the negative
ones, are treated as a uniform background with charge density $-\rho$, in
which particles with positive unit point charges and average particle
density $\rho$ move about \cite{book}, \cite{nothing}.  Since the OCP
background is fixed, charge fluctuations correspond to fluctuations in
particle number, $N_\Lambda$.  $S({\bf r})$ in (\ref{I3}) now corresponds
to the truncated particle-particle correlation function of a system with
average particle density $\rho$ i.e.\ $S({\bf r}) = \{\rho \delta({\bf r})
+ \rho^2[g({\bf r}) - 1]$\} in the conventional liquid theory notation.

Surface area growth of the variance of particle fluctuations is a
conceptually intriguing situation of interest beyond that of equilibrium
Coulomb systems or even beyond statistical mechanics.  Thus, it was proven
in \cite{BECK} that for any system of point particles the variance of 
$N_\Lambda$, averaged over translations and rotations, grows {\it at least}
as fast as  $|\partial \Lambda|$ \cite{BECK}.  Examples of systems having
such a variance are particles arranged on a periodic lattice structure or
those obtained via small distortions of such structures,
\cite{BECK},\cite{KENDAL}.  It was actually proven recently that in one
dimension such growth (which in $d=1$ corresponds to bounded variance)
implies, {\it by itself}, the existence of a periodic component in the
extremal decomposition of any translation invariant measure, a fact already
known for the $d=1$ OCP, using directly methods of equilibrium statistical
mechanics \cite{AGS}.  The existence of such a periodic component implies
that there cannot be good decay of all correlations in a translation
invariant state.  In fact the only example (known to us) of fluctuations in
a point particle system with good mixing (decay of correlation) properties
in $d>1$ is the OCP at high temperatures \cite{I}.  In $d=2$ the
OCP is exactly solvable when the coupling parameter $\Gamma \equiv (kT)^{-1}
= 2$ \cite{JANCO}.  The truncated correlations between groups of
particles separated by a distance $D$ decay in this system in a
super-exponential way, like $\exp[-c D^2]$, with $c$ computed explicitly.
(Interestingly the distribution of particles in the OCP at $\Gamma = 2$ is
the same as that of the (suitably scaled) limiting distribution of
eigenvalues of a random matrix $M$ with entries $M_{ij} \equiv R_{ij} +
iI_{ij}$ in which both $R_{ij}$ and $I_{ij}$ are independent identically
distributed Gaussian random variables \cite{MehCos}.)

In this note we study numerically the dependence of $\hat \kappa$ on the
coupling constant $\Gamma$, for a two-dimensional OCP.  Using a unit of
length proportional to $\rho^{-1/2}$ the characteristic length $\xi_L =
2\hat \kappa/\rho $ depends only on $\Gamma$.  We expect on physical
grounds that $\xi_L(\Gamma)$ will decrease with $\Gamma$ so the question
is: how small can the fluctuations become when $\Gamma \to \infty$?  On the
one hand it is known from numerical studies that the 2d OCP undergoes some
kind of ordering transition to a triangular lattice at $\Gamma = \Gamma_c
\sim 140$ \cite{LW},\cite{Cail}.  On the other hand the exact value of
$\xi_L(\Gamma)$ at $\Gamma=2$ is only about twice the value it would have
if the system was in a rigid triangular lattice and one averaged the
fluctuations over translations and rotations \cite{KENDAL}.  The question
then is how will $\xi_L(\Gamma)$ behave as $\Gamma$ increases towards and
beyond $\Gamma_c$?  For comparison the Debye-H{\"u}ckel length $\xi_{D}$,
which $\xi_L(\Gamma)$ should approach as $\Gamma \to 0$ \cite{BF},
decreases as $\Gamma^{-1/2}$.  If this, or something resembling it, was
also the behavior of $\xi_L(\Gamma)$ we would have a system with
fluctuations much below that of the rigid lattice, which would be
surprising indeed.  It is this question which motivated the investigations
described here.

The results of the simulations are given in Table I and Figure 1.  They
show unambiguously that the decrease in $\hat \kappa(\Gamma)$ saturates for
large $\Gamma$, approaching a value equal (certainly very close) to that of
the triangular lattice.  This is consistent with the intuition that there
is a minimal value of the fluctuation per unit surface area and that this
is achieved for a periodic arrangement of points.  On the other hand it was
recently shown that randomly distributing the position of each particle in
${\Bbb Z}^d$ over a unit cell gives, for $d \gtrsim 350$, smaller
fluctuations than the rigid lattice \cite{Beck}.  The question of the
minimal value of $\hat \kappa$ in different dimensions is still open.

We remark here that we have tried, so far unsuccessfully, to come up with a
scheme for generating translation invariant, mixing measures of particle
distributions in ${\Bbb R}^d$ which would have surface growth of the
variance.  These would be measures on points that do not start with an
equilibrium distribution of the OCP having a finite number of particles in
a box of volume $V$ and then take the thermodynamic limit of $V \nearrow
{\Bbb R}^d$ in an appropriate way to obtain an infinite particle system
with average density $\rho$ \cite{I}.  It is known that the infinite volume
measure of the OCP is not Gibbsian because the probability of large
deviations in the number of particles in a region $\Lambda \in {\Bbb R}^d$
from its average value $\rho|\Lambda|$ behaves like $\exp[-C
|\Lambda|^\gamma]$ with $\gamma > 1$ \cite{jlm}.  One may wonder whether
such behavior necessarily holds for all particle measures on ${\Bbb R}^d$
which have only surface growth of the variance.

\noindent {\bf III.  MODEL AND COMPUTATIONAL DETAILS}

In two dimensions the interaction between two particles of unit charge
separated by a distance $r$ is
\begin{equation}
v(r)=- \ln (r/L) 
\end{equation}
where $L$ is an arbitrary unit length. A convenient unit of length, which
will be used throughout the paper, is the radius of a disk containing one
particle on the average, sometimes referred to as the ``ion-disk radius'',
$a \sim (\pi \rho)^{-{1 \over 2}}$.  The reduced density is then
$\displaystyle \rho = \pi^{-1}$ and a thermodynamic state is uniquely
defined by the coupling constant $\Gamma$.  The difficulties associated
with computer simulations of this system due to the infinite range of the
Coulomb interaction are well known.  They are dealt with here, as in our
previous work, by confining the particles to the surface of a sphere
\cite{Cail}.

For $N$ particles of unit charge moving on the surface of a sphere of
radius $R$ with uniform background density of opposite charge the total
potential energy is taken to be \cite{Cail}
\begin{equation}
V_N = -\frac{1}{2} \sum_{i<j} \ln \Big[ \frac{2R^2}{L^2} (1- {\cc u}_i
\cdot {\cc u}_j) \Big] - \frac{N^2}{4} \Big[1- \ln \frac{4R^2}{L^2} \Big]
\end{equation}
where ${\cc u}_i$ is a radial unit vector locating the position of particle
$i$ on the sphere surface.  This corresponds to the distance between
particles $i$ and $j$ being measured along the chord joining the particles.
In the thermodynamic limit ($ N,R \rightarrow \infty, \rho = N/4 \pi R^2$
constant) the energy differs from that of the planar system by a
contribution of order $O(1/N)$ \cite{Cail}.

Most of our Monte Carlo simulations were performed with $N=1024$ ions (i.\
e.\ $R=\sqrt N/2 = 16$); some at low and high couplings used $N=2048$
($R=22.62$) to check the system size dependence.  Charge fluctuations were
calculated according to Eq.\ (\ref{I4}) which for the 2d OCP takes the form
\begin{equation}
\label{ka}
\hat \kappa = (<N_{\Lambda}^2> - <N_{\Lambda}>^2)/{\cal P}_{\Lambda}
\end{equation}
where $N_{\Lambda}$ is the number of particles in the domain $\Lambda$
  drawn on the surface of the sphere and ${\cal P}_{\Lambda}$ its
  perimeter.  A
  convenient choice for $\Lambda$ is a disk-like shape obtained by the
  intersection of the sphere with a cone of summit at the origin of the
  sphere and aperture $\theta$. To check the independence of the results on
  surface shape additional computations were performed with a
  ``rectangular'' surface obtained by the intersection of the sphere with
  two planes parallel to and symmetric with respect to the equatorial plane
  of the simulation sphere and two parallel planes perpendicular to the
  equatorial plane.

%
%


\noindent {\bf IV.  RESULTS}


\noindent {A.  Charge Fluctuations}

Results for $\hat \kappa $ in the range $\Gamma = 0.01-140$ covering the
whole fluid domain are summarized in Table I and shown in Fig.\ 1.  For
each value of $\Gamma$, particle fluctuations were calculated for domains
of different shape (disk- and rectangular-like) and different area.  For
instance, use of disk-like domains corresponding to $\theta$ =
$81.1^{\circ}$, $72^{\circ}$, $62.4^{\circ}$ and $51.9^{\circ}$, and those
symmetrical with respect to the center of the sphere, gave identical
results (within statistical error) and were therefore averaged over.
Results for rectangular domains (average over 8 domains) and disk-like
domains were found to agree within statistical error (cf.\ Table I).  When
$\Gamma$ is small $\hat \kappa$ decreases rapidly but then saturates at a
value $\hat \kappa=0.042 \pm 0.002$ near $\Gamma=80$. For comparison, the
corresponding value for $\hat \kappa$ for particles on a rigid triangular
lattice is (in our units) $.0404$ while for the square lattice it is
$.0411$ \cite{KENDAL}.

The excellent agreement of the simulation result with the exact value
\cite{AGS} $\hat \kappa= (2 \pi \sqrt{\pi})^{-1}=0.089793$ at $\Gamma = 2$
(cf.\ Table I) shows that both system size and domain dimensions are
sufficiently large for reliable results to be obtained.  In fact system
size dependence is observed only for $\Gamma ^{<}\hskip-1.5ex_{\sim}
0.01$. At these values of $\Gamma$ both the 1024 and 2048
results differ from the Debye-H\"uckel limiting value $(2\pi
\sqrt{2\Gamma})^{-1}$.  This is presumably due to the fact that the
correlation length ($\xi_L \sim \xi_D$) becomes comparable or exceeds the
linear dimension of the domains.

{From} Eq.\ (\ref{I4}) it follows that $\hat \kappa$ can be expressed in 
terms
of the usual pair correlation function $h(r) = g(r) - 1$
\begin{equation}
\label{h}
\hat  \kappa = \displaystyle\-\frac{2}{\pi^2 } \int_0^\infty d {r}\  r^2 h(r)
\end{equation}
where $r$ is in units of $a = (\pi \rho)^{-1/2}$.

An extremely good representation of $\hat \kappa$ in the range $0 \le \Gamma
\le 2$ can be  obtained using in (\ref{h}) an analytical approximation
for the pair correlation function proposed in ref.\  \cite{Pia}.  
\begin{equation}
\label{hx}
 h(x) = \displaystyle -\frac{2}{\Gamma(\mu) } (x \sqrt \mu)^{\mu}
 K_{\mu}(2x \sqrt \mu)
\end{equation}
leading to the simple expression
\begin{equation}
\label{pia1}
\hat \kappa = \displaystyle\frac{1}{2 \pi \sqrt{\pi}} \frac{1}{\sqrt{ \mu}}
\frac{ \Gamma(\mu +3/2)}{\Gamma (\mu +1)}
\end{equation}
where $\displaystyle \mu = \frac{\Gamma}{2-\Gamma} $ and $\Gamma(z)$ and
$K_{\mu}(z)$ are the standard Gamma and Bessel functions, respectively. It
is exact at $\Gamma = 0$ and $\Gamma =2$ and reproduces the simulation data
within statistical error. This is not so surprising since the zeroth,
second and fourth moments of $h(x)$ (Eq.\ (\ref{hx})) are exact
(i.e. perfect screening, Stillinger-Lovett \cite{SL} and compressibility
sum rules are satisfied) and therefore an accurate value for the first
moment Eq.\ (\ref{hx}) can also be expected.

An analogous expression for $\Gamma \ge 2$
\begin{equation}
\label{pia2}
\hat \kappa = \displaystyle\frac{1}{2 \pi \sqrt{\pi}} \frac{1}{\sqrt{ \nu}}
\frac{\Gamma(\nu )}{\Gamma(\nu -1/2)}
\end{equation}
based on Eq.\ (2.27) of ref.\ \cite{Pia}) ($ \nu = \Gamma/(\Gamma - 2)$ 
is less 
satisfactory. Although accurate
near $\Gamma=2$ the subsequent decrease is too slow giving (by analytic
continuation) a value 
$\hat \kappa = \frac{1}{2 \pi^2} = 0.05066$
in the zero temperature limit.

An accurate fit reproducing all the data within statistical error 
is given by
\begin{equation}
\hat \kappa = \displaystyle\frac{0.11253954}{\Gamma^{1/2}} 
\ \displaystyle\frac{1 + a_1 \Gamma^{1/2} + a_2 \Gamma +a_3
\Gamma^{3/2}}{1 + a_4 \Gamma^{1/2} + \Gamma} 
\label{fit}
\end{equation}
 ($a_1=3.9896$, $a_2=1.1211$, $a_3=0.38138$,
$a_4=4.0934$).
This form incorporates the exact Debye-H\"uckel limit 
at low $\Gamma$ and saturates at high values of 
$\Gamma$.


\noindent {B.  Particle distribution} 

As already alluded to in the
Introduction it can be shown that under suitable clustering assumptions,
which are fulfilled for the 2d OCP at small values of $\Gamma$ (and at
$\Gamma = 2$), the probability distribution of
$Q_{\Lambda}/|\partial \Lambda|^{-1/2}$ is Gaussian in the limit $|\Lambda|
\rightarrow \infty$, i.e.\ the probability distribution $P(N_\Lambda)$ of
particles in a domain $\Lambda$ has Gaussian behavior
\cite{GLM},\cite{Lebo}.  To this end a histogram of particle number $N_\Lambda$
was recorded during the simulation and fitted to a Gaussian
%
%
with variance $\sigma^2$. The distributions $P(N_\Lambda)$, calculated in a
disk-like domain with $\theta$ = $81.1^{\circ}$, and its Gaussian fits are
shown in Fig.\ \ref{fig2} for $\Gamma =$ 0.5, 2, 10 and 100.  
We find, for example, after taking an
average over different domains $\Lambda$, that $ \displaystyle \sigma^2
/{\cal P} =$ 0.165, 0.0901, 0.0551, 0.039 for $\Gamma =$ 0.5, 2,10,100,
respectively, in good agreement with the direct calculation (cf.\ Table I).


\noindent {C.  Screening  lengths}

The Debye length $\xi_D$ measures the range of correlations between
pairs of charges in the limit of high temperature \cite{Mart1}, \cite{Bry}.
For low values of $\Gamma$ the asymptotic
behavior of the pair correlation function  exhibits
exponential decay characterized by $\xi_D = 1/ \sqrt{2 \Gamma}$, 
\cite{MehCos},
\cite{Bry}.  This length can be compared with the length $\xi_L$  
typical of the spatial extension of charge fluctuations
defined in (\ref{I4}).  For the density fluctuations in our OCP has
the form 
\begin{equation}
\label{leb}
\hat \kappa = \frac{1}{2} \rho \xi_L =  \frac{1}{2 \pi} \xi_L
\end{equation}
where $\xi_L$ depends only on $\Gamma$ and the factor $1 \over 2$ is chosen
so that  $\xi_L$
agrees  with $\xi_D$ when $\Gamma \to 0$, \cite{BF}.

Table I shows that $\xi_L$ is close to $\xi_D$ for small values of
$\Gamma$, the difference between $\xi_L$ and $\xi_D$ at $\Gamma=2$ is $\sim
10 \%$.  For $\Gamma > 2$ an estimate of the correlation length of the pair
correlation function is more problematic. As shown in Fig.\
\ref{fig4} the pair correlation functions have a damped oscillatory
behavior at long range.  These  oscillations in $h(r)$
can be represented within statistical error by the expression
\begin{equation}
\label{gr}
h(r) \sim  Ae^{- \alpha  r} \cos (\beta r-\gamma)/\sqrt{r}. 
\end{equation}
where $A, \alpha, \gamma$ are fitting parameters. An example of such a fit
is shown in Fig.\ \ref{fig5} for $\Gamma=140$.  Eq.\ (\ref{gr}) can be
obtained by expressing $\tilde{h}(k)$, the Fourier transform of $h(r)$ in
terms of the direct correlation function $\tilde{c}(k)$ according to
$\tilde{h}(k) = \tilde{c}(k)/ (1- \rho \tilde{c}(k))$ and assuming that the
long range behavior of $h(r)$ is driven by the poles in ${(1-\rho
\tilde{c}(k))}^{-1}$ closest to the real axis (``one'' pole approximation
studied, for instance, in \cite{carval} for the 3d OCP).  The parameters
$\alpha$ and $\gamma$ vary little with $\Gamma$. On the other hand
$\alpha$, which fixes the exponential damping of the correlations,
decreases by a factor 3 between $\Gamma =20$ and $140$. Moreover, between
$\Gamma=80$ and $140$, $\alpha$ decreases by a factor 2 whereas $\hat
\kappa$ is nearly constant.  It thus appears that for large $\Gamma$ the
correlations lengths $\xi_L$ and $\xi_D$ are not related directly to the
spatial decay of the envelope of $h(r)$ in the OCP, c.f.\ \cite{BF}. As
mentioned above the exponential damping $\alpha$ is easily obtained from
the ``one pole approximation'' but a more physical explanation of its
origin has still to be found.


\noindent {\bf V.  CONCLUDING REMARKS}

Snapshots showing the arrangement of the ions on the sphere
surface is given in Fig.\ \ref{fig6} for $\Gamma$ = 2 and 140.
These show dramatically how the charges get more uniformly
spaced as the temperature is lowered.  On the other hand  it would be
hard to deduce from looking at the configuration at $\Gamma = 2$ (and even
more so for smaller values of $\Gamma$) that the fluctuations were not
normal:  at $\Gamma = 2$, 
$h(r) = -e^{-\pi \rho r^2}$ so (\ref{I4}) is certainly valid
\cite{JANCO}.  We therefore have to go beyond visual inspection and this can
be done more easily for a multiple component system than for the OCP.  

As was noted in \cite{Lebo} surface area growth of the variance in Coulomb
systems can be interpreted as corresponding to the tendency of charged
particles to form ``bound'' neutral entities. This is most readily
visualized in a two component system of charges $\pm e$ and densities
$\rho_1 = \rho_2 = \rho$.  Suppose now that these charges could be paired
somehow to make neutral dipoles with bond length $D$.  Then the charge
fluctuation in a region $\Lambda$ could be interpreted as resulting from
the boundary $\partial \Lambda$, cutting across some of the ``bonds''
between the charges connecting the ions in these molecules.  We would then
get by standard arguments (assuming sufficient independence between
molecules far apart) a central limit theorem for the fluctuations in
$Q_\Lambda$.  The variance should then be as in (\ref{I4}), i.e.\ of order
CD $e^2 \rho |\partial \Lambda|$ with $CD = \xi_L$.  The constant $C$
measures the effect of correlation between the orientations of the dipoles
cut by $\partial \Lambda$ which is expected to be negative so $C$ should be
less than unity and decrease with $\Gamma$.  The extension $D$ might then
be related to some screening length like $\xi_D$ at high temperatures and
to a ``hard core'' or other minimal distance length at low temperatures and
low densities \cite{STELL}, \cite{BF}, \cite{MEF}.  For the OCP $D \sim a$.

  It is of course not necessary in the above analysis
to insist on pairs of charges, or dipoles, being the basic neutral entity.
We could equally have neutral quadrupoles made up of two positive and two
negative charges or some  hierarchical structure as in the analysis of the
Kosterlitz-Thouless phase in $d=2$ \cite{STELL}, \cite{MEF}.  What is 
necessary to make such
an interpretation of charge fluctuations in $\Lambda$ meaningful is that,
in a typical configuration of the system, the neutral entities be spatially
localized, i.e. not be greatly mixed up with other neutral entities.  This
is of course what happens in insulating materials, be they crystals, gases
or liquids.  Governed by quantum mechanics they consist of tightly bound
neutral atoms or molecules.  The picture in metals is similar in some ways
to that of the OCP.  Apparently enough of this picture remains true
even in classical statistical mechanics of Coulomb systems, i.e.\ perfect
screening, to give the correct behavior of the variance in $Q_\Lambda$.  
%

%
\acknowledgments
%
We would like to thank M. Aizenman, J. Beck, F. Cornu, S. Goldstein,
B. Jancovici and Ph. Martin for helpful discussions and comments.  Computing 
time on the
CRAY C-98 was granted by the Institut de D\'eveloppement et de Ressources
en Informatique (IDRIS).  The Laboratoire de Physique Th\'eorique is
Unit\'e Mixte de Recherche $N^{\circ}$ 8627 du Centre National de la
Recherche Scientifique.  The research of J.L.L. was supported by AFOSR
Grant F49620-98-1-0207, and NSF Grant DMR-9813268.
%
%
\newpage

%
\begin{table}
\caption{Particle number fluctuations  $\hat \kappa =  (<N_\Lambda^2> -
<N_\Lambda>^2)/{\cal P}$ in
the 2d OCP as a function of $ \Gamma$. ${\cal P}$ is the perimeter of the
surface of $\Lambda$. One cycle consists of trial
translations of the $N$ ions. \ $\hat \kappa_{DH}= 1/ (2 \pi \sqrt{2 
\Gamma})$,  $\xi_D = 1 / \sqrt{2 \Gamma}$ and $\xi_L = 2 \pi \hat 
\kappa$.
Length are in units of $(\pi \rho)^{-1/2}$ and  $e = 1$.}
\label{table1}
\begin{tabular}{cccccccc}
$\Gamma$  & $N$  &  
cycles & $\hat \kappa$  & $\hat \kappa_{DH}$ & Eqs.\ 13 and 14 & $\xi_L$ &
$\xi_D$   \\ \hline  
0.01 & 1024 &100000 & $1.035 \pm 0.008$ &     &  &    \\
0.01 & 2048 &100000 & $1.080 \pm 0.007$ & 1.1254 & 1.1260 &6.80 & 7.071 \\
0.05 & 1024 &100000 & $0.499 \pm 0.001$ &  0.5033 & 0.5047 &3.135 & 3.162  \\
0.10 & 1024 &200000 & $0.356 \pm 0.001$  & 0.3559 &0.3579 & 2.237 & 2.236\\
0.5 & 1024 & 300000 & $0.1638 \pm 0.0006$  & 0.1592 & 0.1638 & 1.029&1.00 \\
1 & 1024 & 200000 & $0.1197 \pm 0.0004$  & 0.1125 &0.1194 &0.752&0.707 \\
$2{}^a $ & 1024 & 100000 & $0.0897 \pm 0.0004$ & 0.0796 &0.0898 & 
0.564&0.50\\
4  & 1024 & 200000 & $0.0701 \pm 0.0002$  & 0.0563 &0.0716&0.440&0.354 \\ 
$4{}^b$& 1024 & 100000& $0.0696 \pm 0.0003$ & & &  \\
10  & 1024 & 100000 & $0.0548  \pm 0.0005$  &  &0.0594 & 0.344  \\
20  & 1024 & 100000 & $0.0487  \pm 0.0005$ & & & 0.306  \\
40  & 1024 & 200000 & $0.0450  \pm 0.0012$ & & & 0.283  \\
$40{}^b$ & 1024 & 200000 & $0.0444 \pm 0.0020$  \\
60  & 1024 & 200000 & $0.0435 \pm 0.0009$ & & & 0.273   \\
80  & 1024 & 200000 & $0.0428 \pm 0.0012$ & & & 0.269   \\
$80{}^b$ & 1024 & 100000 & $0.0423 \pm 0.0020$  \\
100  & 1024 & 400000 & $0.0427  \pm 0.0007$   \\
100  & 2048 & 90000 & $0.0431  \pm 0.0012$  & & & 0.271  \\
120  & 1024 & 200000 & $0.0416  \pm 0.0007$ & & & 0.261   \\
140  & 1024 & 400000 & $0.0417 \pm 0.0010$ &  &  \\
140  & 2048 & 100000 & $0.0421 \pm 0.0026$ & &0.051 & 0.264  \\
\end{tabular}
${}^a$ \quad the exact result is $0.089793$ \cite{JANCO}, \quad ${}^b$  results for ``rectangular'' surfaces.
\end{table}

\begin{table}
\caption{Parameter $\alpha$ of the asymptotic decay of the pair correlation 
function $h(x)  \approx Ae^{- \alpha r} \cos (\lambda
r-\gamma)/\sqrt{r}$ 
(where $r$ is in units of $a$  the ion-disk radius).}
\label{table2}
\begin{tabular}{cccccccc}
$\Gamma$  & 20 & 40 & 60 & 80 & 100 & 120 & 140      \\ \hline 
$\alpha$  & 0.769 &0.549&0.404&0.326&0.278&0.225&0.194  \\
\end{tabular}
\end{table}
%
%

%
\begin{figure}
\caption{Variation of $\hat \kappa$ with coupling $\Gamma$\hfill}
\label{fig1}
\end{figure}
%

%

\begin{figure}
\caption{Particle number distribution $P(N)$ for $\Gamma=$ 0.05, 0.5, 2,
10 and 100 (from bottom to top). The filled circles are the Monte Carlo
and the lines represent the Gaussian fits.}
\label{fig2}
\end{figure}

\begin{figure}
\caption{Pair distribution function $g(r)$ for $\Gamma=$ 10, 20, 40, 60,
80, 100, 120 and 140 (with increasing peak heights).} 
\label{fig4}
\end{figure}
\begin{figure}
\caption{Fit of the long range oscillations of the pair correlation
function $h(r)=g(r)-1$ for $\Gamma=$ 140 by means of the functional form
Eq.\ \ref{gr}.} 
\label{fig5}
\end{figure}
\begin{figure}
\caption{Snapshots of  particle configurations on the sphere surface at
$\Gamma=$ 2, 140. Only the ions on the visible part of the surface are
 shown.  (1024 particles were used for $\Gamma = 2$ and 2048 for
$\Gamma = 140$.)} 
\label{fig6}
\end{figure}

\newpage
%

%
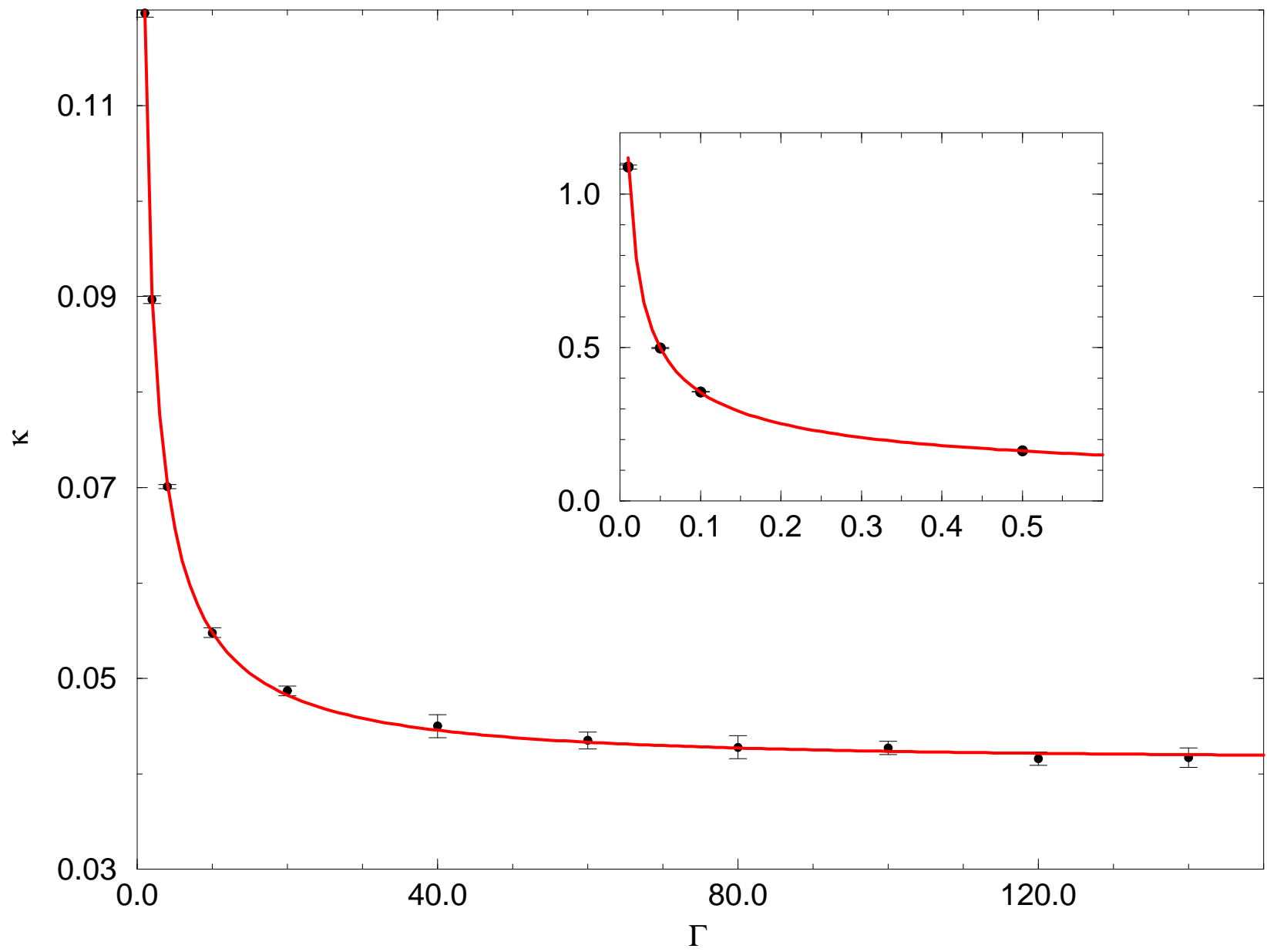
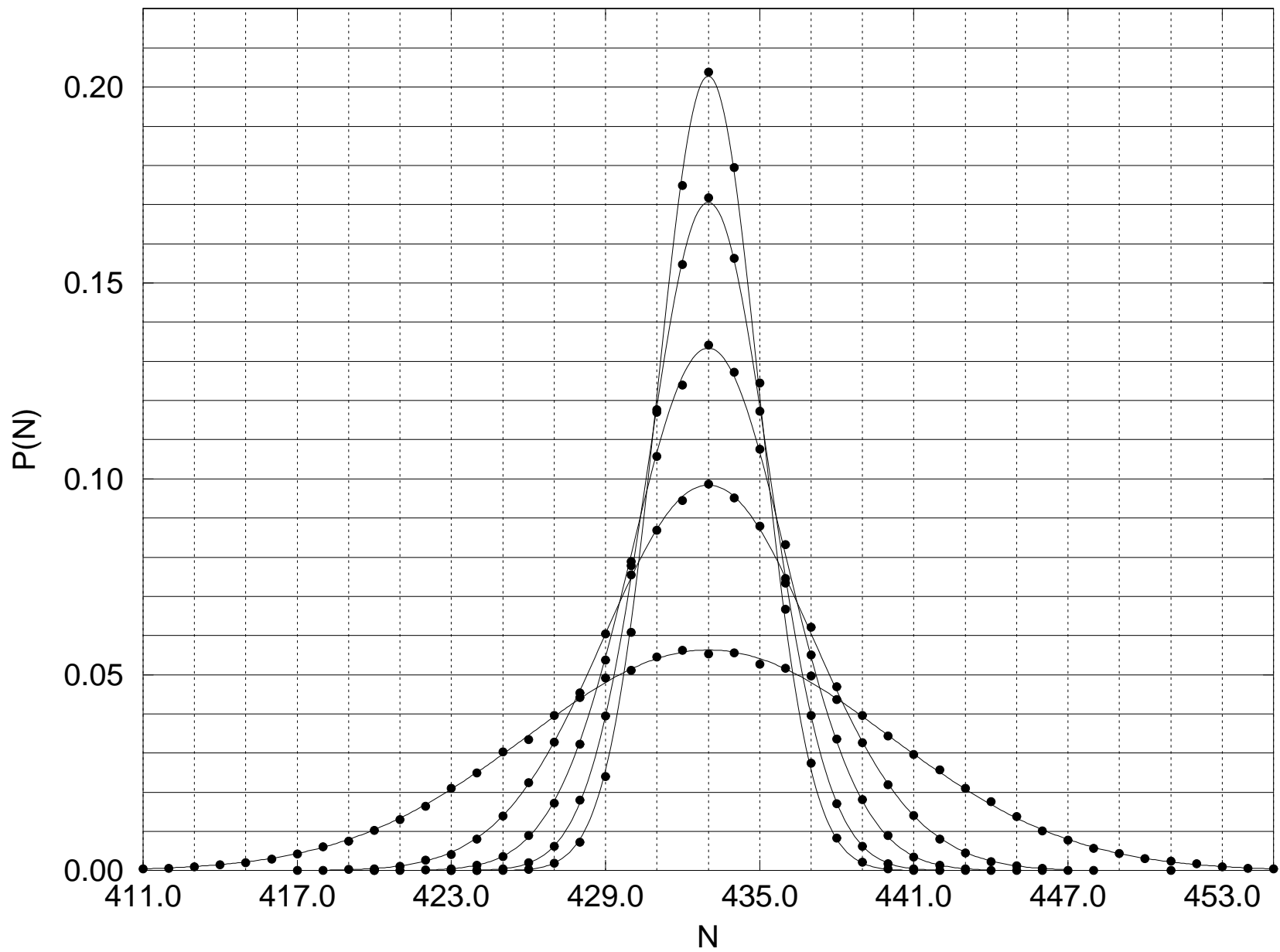
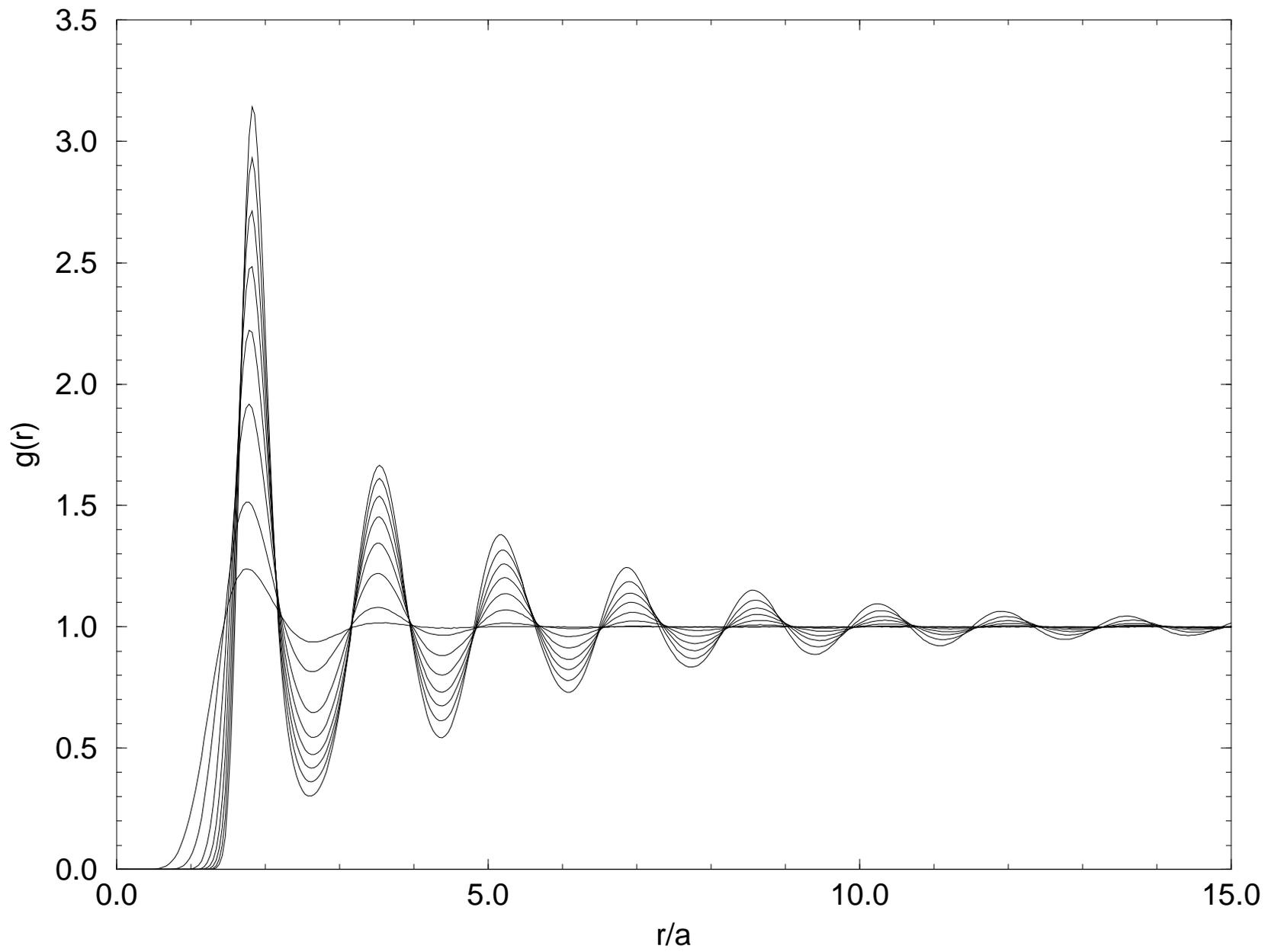
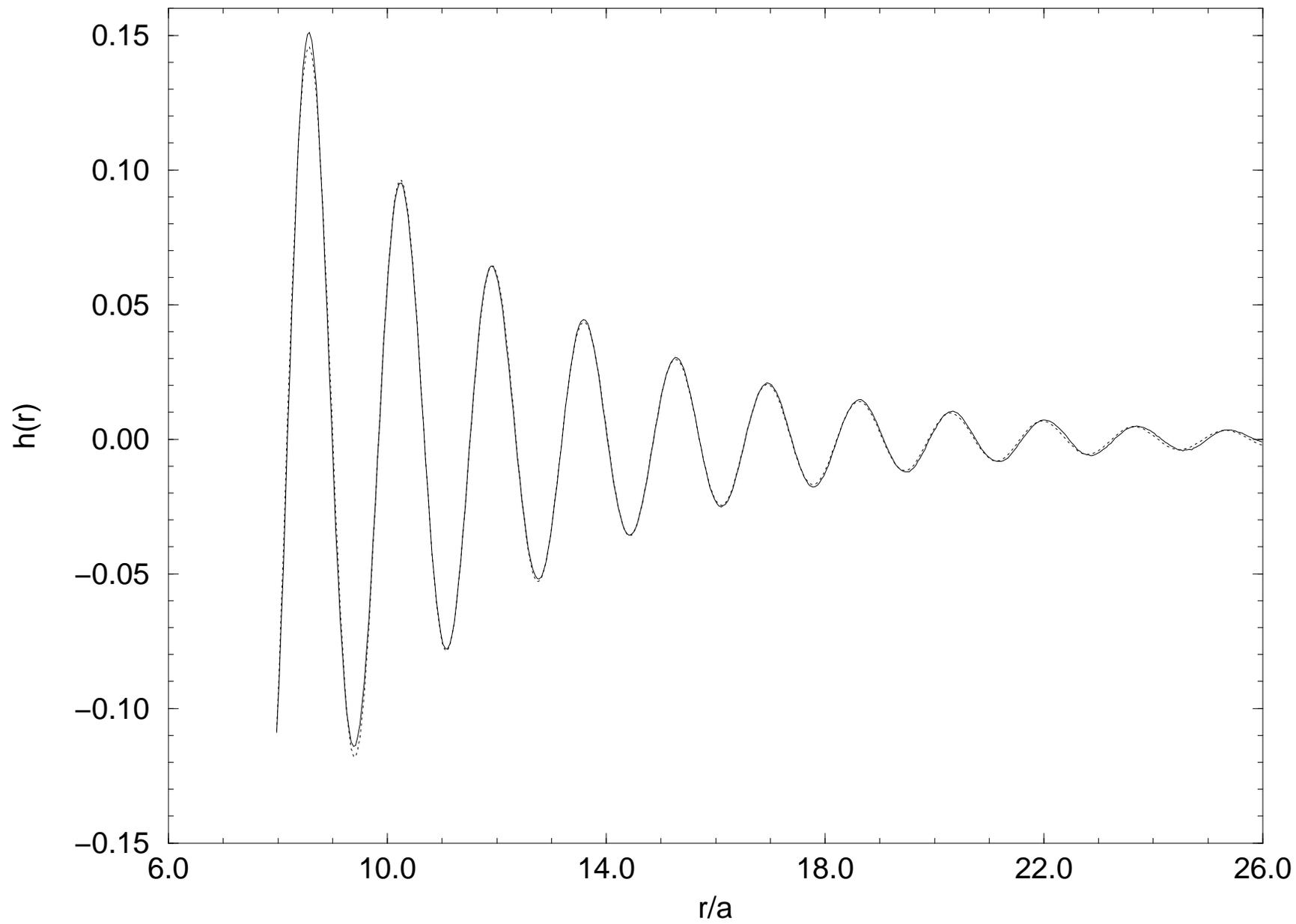
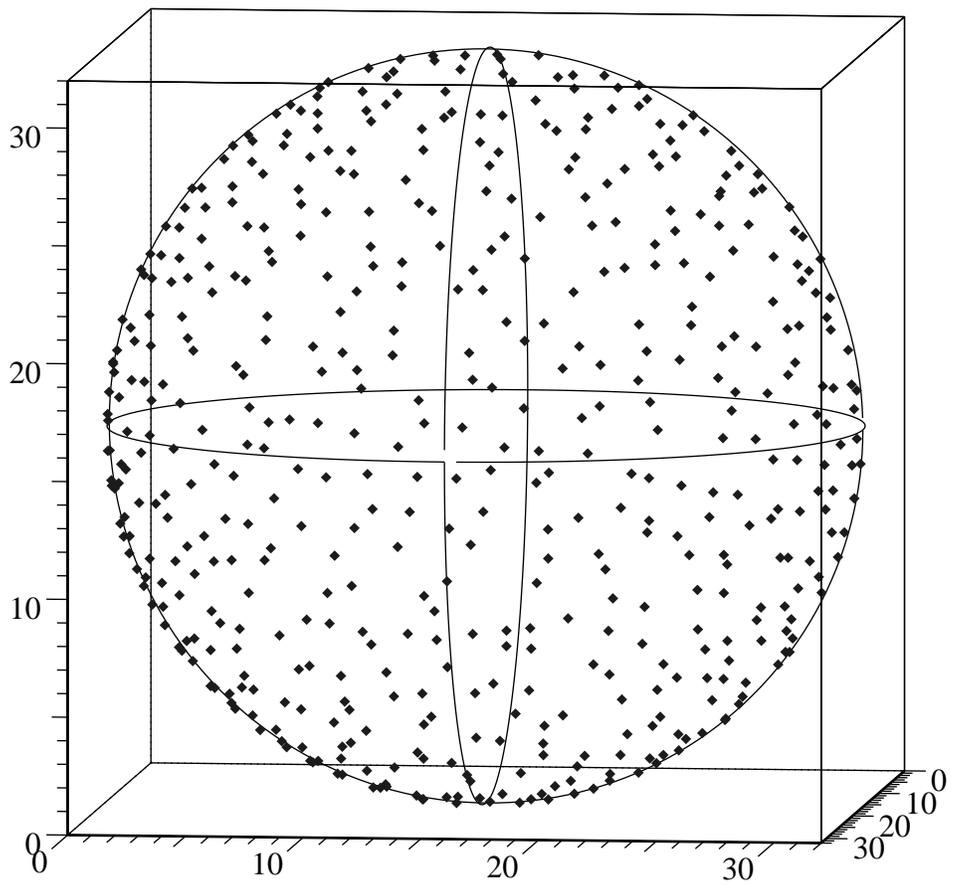
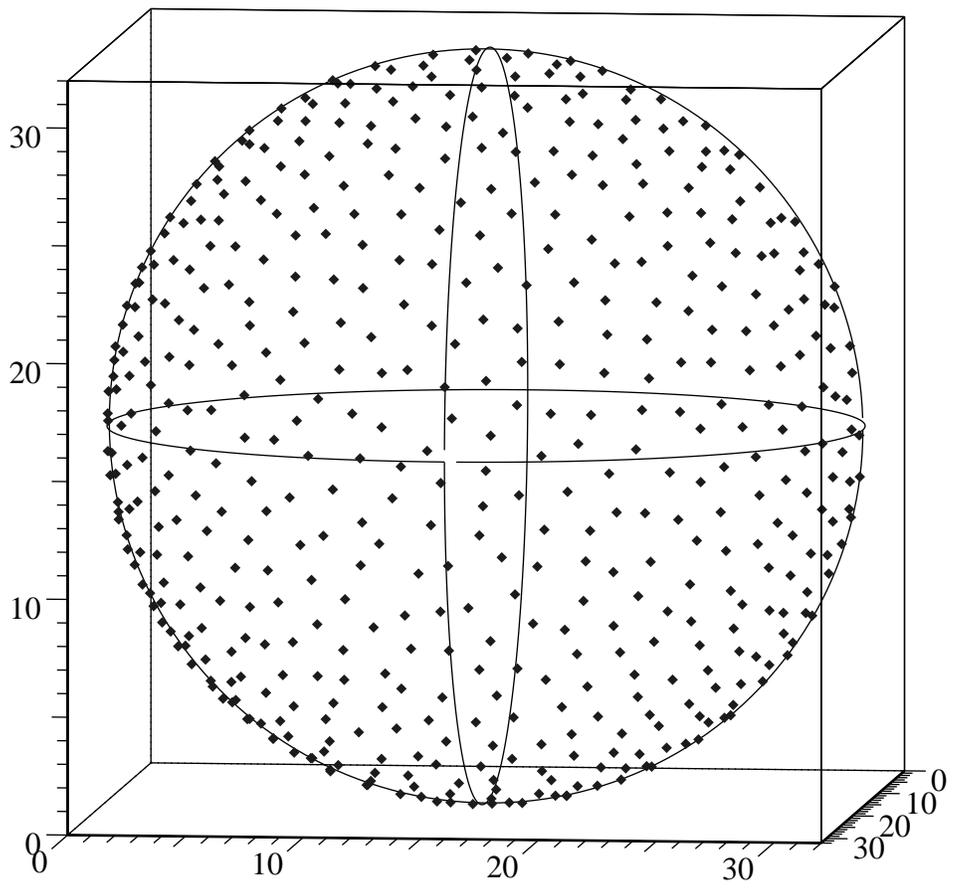
\end{document}